\documentclass[pre,twocolumn,showpacs]{revtex4}
\usepackage{epsfig,amssymb,graphicx}

\begin{document}

\title{The Shape and Stability of a Viscous Thread.}
\author{Sergey Senchenko$^{1,2}$, Tomas Bohr$^{1,3}$.}
\affiliation{$^{1}$Physics Dept., Danish Technical University, DK-2800
Lyngby, Denmark}
\email{senchen@fysik.dtu.dk}
\affiliation{$^{2}$Optics and Fluid Dynamics Department, 
Ris\o ~National Laboratory, DK-4000 Roskilde, Denmark}


\date{\today}
\begin{abstract}
When a viscous fluid, like oil or syrup, streams from a small orifice
and falls freely under gravity,
it forms a long slender thread, which can be maintained in a stable,
stationary state with 
lengths up to several meters. 
We shall discuss the shape of such liquid threads and their surprising
stability. 
It turns out that the strong advection of the falling fluid can almost
outrun the Rayleigh-Plateau
instability. Even for a very viscous fluid like sirup or silicone oil,
the asymptotic shape and stability is independent of viscosity
 and small perturbations grow with time as
$\exp({{\rm C} \,t^{\frac{1}{4}}})$, where the constant is independent of
viscosity. The corresponding spatial growth has the form
$\exp({(z/L)^{\frac{1}{8}}})$, where $z$ is the down stream distance and $L
\sim  Q^2 \sigma^{-2} g$ and where $\sigma$ is the surface tension,
 $g$ is the gravity and $Q$ is the flux. However, the value of viscosity
determines the break-up length of a thread $L_{\nu} \sim \nu^{1/4}$
and thus the possibility of observing the $\exp({{\rm C} \,t^{\frac{1}{4}}})$ 
type
asymptotics.
\end{abstract}
\maketitle
\section{Introduction}
When honey or sirup is poured from an outlet, one easily generates very
long threads of flowing fluid of surprising beauty and stability.
A uniform column of fluid is unstable due to surface tension effects -
the famous Rayleigh-Plateau instability \cite{chandra}. 
Viscosity diminishes the strength, but does not remove the instability,
and thus the observation of stable falling viscous threads of, say,
two meters is surprising. In the present paper we shall discuss the
shape and stability of such falling viscous 
jets or threads. We should note from the outset that we are confining
our attention to Newtonian fluids (e.g. sirup or Silicone oil).

\par
Our starting point is the lubrication approximation (see. e.g.
\cite{E97}), which only takes
 into account the leading order dependence of the velocity
field on the radial variable and of which we give a short derivation in
section II. 
In section  III we study the stationary solutions, and in particular
their asymptotic forms. 
The final asymptotics (for large downstream distance $z$) is always
governed solely by gravity, as in a free fall.
Then we proceed with linear stability analysis (section IV-IV). 
After a recapitulation of  the classical Rayleigh-Plateau instability in
the lubrication approximation 
in the absence of gravity,
we then study the full linear stability problem of a falling thread using 
an Eulerian description in the comoving frame. 
We consider first the inviscid regime, which determines the asymptotics 
for large times if it exist at all. We then asses the importance of 
viscosity and find that it determines the opposite asymptotics
of spatially growing modes at short times. Small viscosity thus means that 
the flow breaks up without ever reaching the inviscid asymptotic state.

\section{Derivation of the Model.}
To derive our model we use the lubrication approximation, and our 
derivation is thus very close to  the one given e.g by Eggers \cite{E97}. 
Since we wish to include gravity our equations are scaled differently, 
however.
The velocity field is assumed to be axisymmetrical, and it is
convenient to use cylindrical coordinates. 
 We thus assume that the velocity field has the form: $\vec v=u
\vec{e}_r +w \vec{e}_z $, where $r$ is the
 radial coordinate and $z$ is the vertical coordinate, measured positive
downwards. We assume that the velocity
 field has no azimuthal component.
The Navier-Stokes equation and continuity equation \cite{LL}
\begin{eqnarray}
u_{t}+uu_{r}+wu_{z}&=
&-p_{r}/\rho+\nu\left ((ru_{r})_{r}/r + u_{zz} -u/r^2\right) \nonumber \\
w_{t}+uw_{r}+ww_{z}&=
&-p_{r}/\rho+g +\nu\left ((rw_{r})_{r}/r + w_{zz}\right) \label{ns}\\
(ru)_{r}+w_{z}&=&0\nonumber
\end{eqnarray} 
and we assume that the fluid is confined to a thin axially symmetric
thread with a free surface at
$r=h(z,t)$.
We expand pressure and velocity fields in power series in $r$, and 
assume that the expansion parameter $r$ is  
(asymptotically) small with respect to
vertical coordinate $z$ at a given cross-section of the fluid thread, 
i.e. $r/z \rightarrow 0, z \rightarrow \infty$:
\begin{eqnarray}
w&=&w_{0}(z,t)+w_{2}(z,t) r^2 +\ldots\nonumber\\
u&=&-w_{0z}(z,t)r/2-w_{2z}(z,t)r^3 /4+\ldots\label{ex} \\
p&=&p_{0}(z,t)+p_{2}(z,t)r^2 +\ldots\nonumber
\end{eqnarray}
Here the expression for $u$ guarantees 
that the velocity field is divergence-free.
\newline
Inserting this expansion into the Navier-Stokes equation gives to
leading order
\begin{eqnarray}
w_{0t}+w_{0}w_{0z}=-p_{0z}/\rho+g+\nu(w_{0zz}+4w_{2}) \label{m_eq}
\end{eqnarray}
To close the equation, we need to express $w_2$ and $p_{0}$ in terms of
$w_0$ by using the dynamic boundary condition.
\begin{eqnarray}
\hat \sigma \vec n=-\alpha \kappa \vec n \label{bc}
\end{eqnarray}
where $\hat \sigma$ is a stress tensor, $\kappa$ is a mean curvature
of the surface, $\alpha$ is the coefficient of surface tension
and $\vec n$ is a unit normal vector, pointing into
the fluid. In terms of the fluid surface 
$r=h(z,t)$, the normal vector is $\vec n=(n_{r},n_{z})=(-1,h_{z})/ 
\sqrt{1+h_{z}^2}$.
\newline
The only nonzero components of stress tensor are (see e.g. \cite{LL}):
\begin{eqnarray}
\sigma_{rr}&=&-p+2\nu\rho u_{r} = -p_{0}-\nu\rho w_{0z}+... \nonumber\\
\sigma_{zz}&=&-p+2\nu\rho w_{z} = -p_{0}-\nu\rho w_{0z}+...
\label{stt} \\
\sigma_{rz}&=&\nu\rho(u_{z}+w_{r}) = (2w_{2}-w_{0zz}/2)r+... \nonumber
\end{eqnarray}
and inserting into (\ref{bc}) 
\begin{eqnarray}
\sigma_{rr}n_{r}+\sigma_{rz}n_{z}=-\alpha\kappa n_{r} \nonumber \\
\sigma_{zr}n_{r}+\sigma_{zz}n_{z}=-\alpha\kappa n_{z}  \label{sys}
\end{eqnarray}
gives (canceling the common multiplier $1/ \sqrt{1+h_{z}^2})$
\begin{eqnarray}
p_{0}+\nu\rho w_{0z}+\nu\rho (2w_{2}-w_{0zz}/2)h h_{z}&=&\alpha \kappa
\label{sys1} \\
\nu\rho (2w_{2}-w_{0zz}/2)h-(-p_{0}+2\nu\rho w_{0z})h_{z}&
=&\alpha \kappa h_{z}
\end{eqnarray}
Neglecting, again for a thin thread, the nonlinear term $h h_z$ we obtain
\begin{eqnarray}
p_{0} &=& \alpha \kappa - \nu \rho w_{0z} \label{pr} \\
w_{2} &= & {\frac{3}{2}} w_{0z}r_{z}/r+{\frac{1}{4}} w_{0zz} \label{w2}
\end{eqnarray}
Using these expressions for $p_0$ and $w_2$ in (\ref{m_eq}) we get:
\begin{eqnarray}
w_{0t}+w_{0}w_{0z}=-\frac{\alpha}{\rho} 
\kappa_{0z}+g+\nu\left(3w_{0zz}+6w_{0z}\frac{h_{z}}{h}\right)
 \label{m_eq1}
\end{eqnarray}
The kinematic boundary condition leads to the conservation law for the
cross-section of the thread:
\begin{equation}
(h^{2})_{t}+(w_{0}h^{2})_{z}=0,
\end{equation}
The curvature is:
\begin{eqnarray}
\kappa= 
(\sqrt{1+h_{z}^2})^{-1}\left(\frac{1}{h}-\frac{h_{zz}}{1+h_{z}^2}\right)
\label{cur_c}
\end{eqnarray}
but since we are here interested in asymptotic properties of thin
threads, we neglect the curvature 
in the $(r,z)$ plane compared to the one around the axis of the thread,
and assume that $h_z \ll 1$.
 Thus we shall, throughout the paper use the approximation:
\begin{eqnarray}
 \kappa \approx  \frac{1}{h}
\label{cur_t}
\end{eqnarray}
Thus the our model has the following form (using $\sigma=\alpha/\rho$):
\begin{eqnarray}
(h^{2})_{t}+(w_{0}h^{2})_{z}&=&0, \label{system_1} \\
w_{0t}+w_{0}w_{0z}&=&-\sigma \left(\frac{1}{h}\right)_{z}+g+
3\nu\frac{(w_{0z}h^2)_{z}}{h^2}\label{system_2}
\end{eqnarray}
We now introduce dimensionless variables through
\begin{eqnarray}
z \rightarrow \alpha z, \ \ \ t \rightarrow \beta t \ \ \ 
h \rightarrow \alpha h, \ \ \ w_{0} \rightarrow \frac{\alpha}{\beta}v
\label{subst}
\end{eqnarray}
where $\alpha$ and $\beta$ are dimensional coefficients.
Thus (\ref{system_2}) 
acquires the following form:
\begin{eqnarray}
v_{t}+v v_{z}&=&\frac{\beta^{2}}{\alpha}g
-\frac{\beta^{2}}{\alpha^{3}}\sigma \left(\frac{1}{h}\right)_{z}+
3\nu\frac{\beta}{\alpha^{2}} \frac{(v_{z} h^{2})_{z}}{h^{2}}
\label{main2}
\end{eqnarray}
whereas (\ref{system_1}) preserves its form
since it is homogeneous in space and time variables. 
\par
We choose $\alpha$ and $\beta$ such that the two first coefficients on
the RHS of (\ref{main2}) is unity, i.e. :
\begin{eqnarray}
\alpha=\sigma^{\frac{1}{2}}g^{-\frac{1}{2}}, \  \  \ 
\beta=\sigma^{\frac{1}{4}}g^{-\frac{3}{4}}
\label{coeff2}
\end{eqnarray}
This allows us to consider both viscid and inviscid cases by means of
the last coefficient $ \gamma = 3\nu{\beta}/{\alpha^{2}} = 3\nu
\sigma^{-3/4} g^{-1/4}$.
Note that this choice of rescaling means that  lengths are measured in
units of the capillary length
$l_{c}= \sigma^{\frac{1}{2}}g^{-\frac{1}{2}} =  \alpha$. 
Thus the non-dimensionlized model has the following form:
\begin{eqnarray}
(h^{2})_{t}+(vh^{2})_{z}&=&0 \label{ourmodel_1} \\
v_{t}+vv_{z}&=&-\left(\frac{1}{h}\right)_{z}+1+
\gamma\frac{(v_{z}h^2)_{z}}{h^2}\label{ourmodel_2}
\end{eqnarray}

Typical values for $\gamma$ are $\gamma_{sirup} \approx 100$ (with
similar values for heavy silicone oils), 
$\gamma_{glycerol}\approx 0.4$ and $\gamma_{water} \approx 0.004$.

Solution
\section{Stationary Solutions.}
The shape of a stationary thread has been studied by several authors (see
\cite{QJ}-\cite{G&S}), but since the results are somewhat scattered and 
incomplete,
we have found it important describe the stationary states in some detail.
For stationary solutions the non-dimensional flux: $q=h^{2}v$ is
constant and we end up with the following equation for 
velocity field only:
\begin{eqnarray}
vv_{z}=1-\frac{v_{z}}{2\sqrt{qv}}+\gamma v_{zz}-\gamma
\frac{v_{z}^{2}}{v}. \label{stat}
\end{eqnarray}
It is possible to remove dependence on $q$ by an appropriate rescaling,
but we prefer to keep it
since the total flux is the only parameter of the model that can easily
be changed in a typical experiment. The flux $q$ is by the scaling
(\ref{coeff2}) related to the
physical flux $Q$ as
\begin{eqnarray}
q= \alpha^{-3}\beta Q/\pi=\sigma^{-5/4}g^{3/4}Q/\pi. \label{phflux}
\end{eqnarray} 
made
\par
\begin{center}
\begin{figure}
\begin{center}
 \includegraphics[width=80mm]{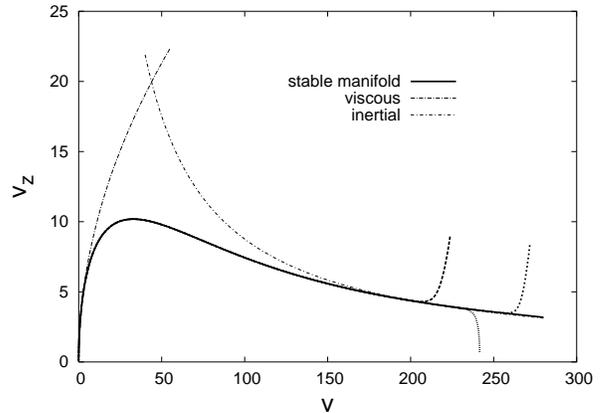}
  \end{center}
\caption{The phase plane for the equation (\ref{stat}). It is seen that,
upon backward integration, trajectories quickly converge to a
well-defined "unstable manifold" (full line) for the fixed point
$(v,v_z)=(0,0)$. The asymptotic solution for large $z$, $v \sim
\sqrt{z}$, 
is shown dot-dashed and is governed by inertia and gravity.
The asymptotic solution for small $z$, $v \sim z^2$, is shown dotted and
is obtained by neglecting inertia. }
\label{ph_plane} 
\end{figure}
\end{center}

When (\ref{stat}) is solved forward in $z$, i.e. as an "initial value
problem", the typical solutions will diverge for large $z$. This can be
circumvented by integrating backwards noting that the fixed point
$(v,v_z) = (0,0)$ has a well-defined unstable manifold (separating
solutions that diverge to plus or minus infinity), which upon backward
integration becomes a stable manifold. In 
Fig.~\ref{ph_plane} we show typical phase space trajectories found by
solving (\ref{stat}) numerically by means of a fourth-order Runge-Kutta
method, starting from "initial conditions" $(v_{0}, v_{z0})$ at large
$z$ and integrating backwards.
It is seen that the dependence on the particular choice of
downflow conditions is very week since any phase trajectory quickly
converges to the well defined stable manifold. Thus, even for a thread
of moderate 
length the shape is uniquely determined irrespective of the precise
downstream conditions, just as we would expect.

The asymptotic behaviour of the solution as $z \rightarrow \infty$ is
easily seen to be controlled by only the two first terms in
(\ref{stat}), i.e. 
\begin{equation}
vv_{z}=1
\end{equation}
giving 
\begin{eqnarray}
v&=& \sqrt{2z}  \label{v_inert}\\
h&=& \sqrt{\frac{q}{v}} = q^{1/2} (2z)^{-1/4} \label{h_inert}
\end{eqnarray}
This asymptotic solution is shown by the dot-dashed curve in
Fig.~\ref{ph_plane} (marked "inertial"). 

The behaviour of the unstable manifold near the fixed point $(v,v_z) =
(0,0)$ can be found by expanding in $z$. Clearly $v = C z^2 + O(z^3)$
for the RHS of (\ref{stat}) to remain finite as $z \rightarrow 0$.
Inserting this expression into (\ref{stat}), we see that the inertial
term $v v_z$ can be neglected, since it contributes only as $z^3$,
whereas all other terms contribute with $z^0$-terms, and we find
\begin{equation}
1 -\sqrt{C/q} - 2 \gamma C = 0
\end{equation}
with the (positive) solution 
\begin{eqnarray}
\label{C}
C=\frac{1+4\gamma q -\sqrt{1+8\gamma q}}{8\gamma^{2}q}.
\end{eqnarray}
With this choice of $C$ the solution
\begin{eqnarray}
v&=& C z^2 \label{v_visc}\\
h&=& \sqrt{\frac{q}{C}}z^{-1} \label{h_visc}
\end{eqnarray} 
is in fact an exact solution to (\ref{stat}), when the inertial term $v
v_z$ is neglected. This $v(z)$ is shown by the dotted curve in
Fig.~\ref{ph_plane} (marked "viscid"). 

\begin{center}
\begin{figure}
\begin{center}
\includegraphics[width=80mm]{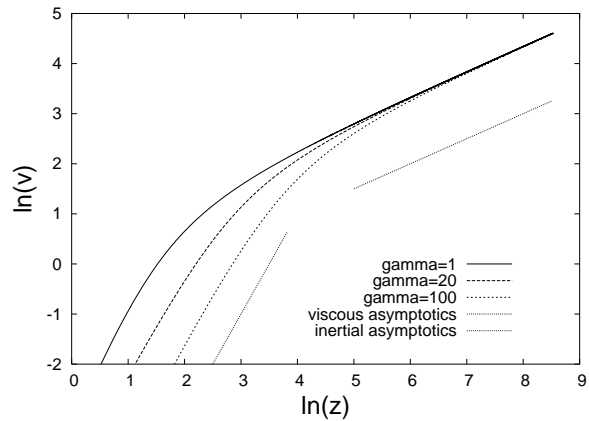}
  \end{center}
\caption{Plot of the numerical solution of (\ref{stat}) for
different values of  $\gamma$}
\label{log-log} 
\end{figure}
\end{center}

For a very viscous fluid, where $\gamma q \gg 1$, the coefficient is
\begin{eqnarray}
C \approx {\frac{1}{2 \gamma}}
\label{large_gamma}
\end{eqnarray}
The crossover between the viscid and inertial solutions is roughly given
by the value $z^*$ where
they become equal, i.e.
\begin{equation}
z^* \approx {\rm const} ({\frac{\sqrt{2}}{C}})^{2/3} \sim \gamma^{2/3} 
\label{z*}
\end{equation}
On Fig.~\ref{log-log} we plot $v(z)$ for various values of $\gamma$.

\section{Stability properties.}
\subsection{Stability of a fluid cylinder in the lubrication 
approximation}
Let us quickly go through the stability in this case using the dimensional 
equations (\ref{system_1})-(\ref{system_2}) instead of the dimensionless 
(\ref{ourmodel_1})-(\ref{ourmodel_1}) since we want to take the limit 
$g=0$.
In the absence of gravity term (the constant 1 on the RHS) where the
stationary state is a cylinder moving with constant velocity. This is
the classical
Rayleigh-Plateau instability in the long wave length approximation  
\cite{E97}. We
thus assume
\begin{eqnarray}
v&=&v_{0}+\tilde v(z,t),\nonumber \\
h&=&h_{0}+\tilde h(z,t).\nonumber
\end{eqnarray}
and obtain the linearized system
\begin{eqnarray}
\tilde v_{t}+ v_{0} \tilde v_{z}&=& \sigma \frac{\tilde h_{z}}{h_{0}^{2}}
+\nu \tilde v_{zz}, \label{evh} \\
\tilde h_{t}
+v_{0}\tilde h_{z}&=&-\frac{1}{2}h_{0}\tilde v_{z}. \nonumber
\end{eqnarray} 
It is convenient to go to the comoving frame 
\begin{eqnarray}
y=z-v_{0}t,\ \ \ \tau=t.
\end{eqnarray}
where
\begin{eqnarray}
\tilde v_{\tau}&=& \sigma \frac{\tilde h_{y}}{h_{0}^2}+
\nu \tilde v_{yy},\label{lin} \\
 \tilde h_{\tau}
&=&-\frac{1}{2}h_{0}\tilde v_{y}.
\end{eqnarray} 
Transforming as usual to Fourier modes as
\begin{eqnarray}
(\tilde v,\tilde h) = (C_{1},C_{2})exp(ix y + s\tau), \label{plane}
\end{eqnarray}
leads to  the dispersion relation:
\begin{eqnarray}
s_{\pm}=\frac{1}{2}\left[-\nu x^2 \pm \sqrt{\frac{2 \sigma
x^2}{h_{0}}+\nu^2 x^4} \right]
\label{rp_disp}
\end{eqnarray}
which, within the long-wave region $x \ll h_{0}$
coincides with the well-known results
for the classical Rayleigh-Plateau instability \cite{chandra}.

In the inviscid case $\nu=0$, we get (for the unstable mode with
positive $s$):
\begin{equation}
s = x \sqrt{\frac{\sigma}{2h_{0}}}. \label{invisc_s}
\end{equation}

\subsection{Stability of the inviscid thread solution}
We now study directly the stability of the stationary states of 
(\ref{ourmodel_1})-(\ref{ourmodel_2})
in the limit of vanishing viscosity, i. e.
\begin{eqnarray}
h^{2}_{t}+(vh^{2})_{z}&=&0, \label{inv_eul} \\
v_{t}+vv_{z}&=&-\left(\frac{1}{h}\right)_{z}+1.
\end{eqnarray}
and we linearize around the stationary solution $(v_0(z),h_0(z))$
(\ref{v_inert})-(\ref{h_inert}) as
\begin{eqnarray}
v&=&v_{0}(1+a) \label{raz}\\
h&=&h_{0}(1+b) \nonumber
\end{eqnarray}
to obtain the linear system:
\begin{eqnarray}
a_{t}+v_{0}a_{z}+2v_{0z}a&=&q^{-1/2}v_{0}^{-1/2}b_{z} \nonumber\\
&+&\frac{1}{2}q^{-1/2}v_{0}^{-3/2}v_{0z}b \label{lin1} \\
b_{t}+v_{0}b_{z}+\frac{v_{0}}{2}a_{z}&=&0 \label{lin2}
\end{eqnarray}
To get rid of the advection term, we introduce the stretched spatial
variable $y$ as
\begin{equation}
y = \int {\frac{dz}{v_0(z)}} \label{dif_var}
\end{equation}
so that $v_{0}(z)\partial_{z}=\partial_{y}$. We also define the function
$W(y)$ as
\begin{equation}
W(y) = v_{0}(z(y)) \label{W}
\end{equation}
and these definitions transform (\ref{lin1}) - (\ref{lin2}) into
\begin{eqnarray}
a_{t}+a_{y}+2W^{-1}W_{y}a&=&q^{-1/2}W^{-3/2}b_{y}\nonumber\\
&+&\frac{1}{2}q^{-1/2}W^{-5/2}W_{y}b  \label{lin1a}\\
b_{t}+b_{y}+\frac{a_{y}}{2}&=&0 \label{lin2a}
\end{eqnarray}
For the inertial stationary solution $v_0(z) = \sqrt{2 z}$ we have 
explicitly
\begin{eqnarray}
\label{str}
z=\frac{y^2}{2}, \ \ \ W(y)=y
\end{eqnarray}
and we finally transform (\ref{lin1a}) - (\ref{lin2a}) into the comoving
frame of reference by
\begin{eqnarray}
y=x+t, \ \ \ t=t \label{comov}
\end{eqnarray} 
to obtain
\begin{eqnarray}
b_{t}+\frac{a_{x}}{2}&=&0  \label{lin1b}\\
a_{t}+2 a (x+t)^{-1}&=&q^{-1/2}(x+t)^{-3/2}b_{x}\nonumber\\
&+&\frac{1}{2}q^{-1/2}(x+t)^{-5/2}b 
\end{eqnarray}
We now Fourier-transfom in $x$, assuming that the asymptotic behaviour
will not be influenced by the 
slow algebraic variation with $x$ in the denominators as long as $t \gg 
x$, an assumption which will be verified in the Appendix. Thus we find, in
terms of the Fourier transforms $\tilde a(k,t)$ and $\tilde b(k,t)$,
\begin{eqnarray}
\tilde a &=& \frac{2i}{k} \tilde b_{t} \nonumber\\
{\tilde a}_{t} & \approx&  - 2 {\tilde a} t^{-1} + i k q^{-1/2} t^{-3/2}
{\tilde b}
+\frac{1}{2}q^{-1/2} t^{-5/2} {\tilde b}
\label{lin2b}
\end{eqnarray}
Making finally the substitution
\begin{eqnarray}
\tilde b= t^{-1}B
\end{eqnarray}
we find, retaining only the dominant term as $t \rightarrow \infty$
\begin{eqnarray}
B_{tt}= \frac{k^{2}q^{-1/2}}{2}t^{-3/2} B
\end{eqnarray}
The WKB ansatz
\begin{equation}
B (k, t) = B_0 \exp (\int^{t} \Phi (k, t') d t')
\end{equation}
gives
\begin{eqnarray}
B_{+}(t)=\exp(2\sqrt{2}kq^{-1/4}t^{1/4}) \label{one_q1}
\end{eqnarray}
Thus the typical instability time is $ t_c \sim q$ and using
(\ref{phflux}) and {\ref{coeff2}) allows us to estimate
 the typical dimensional instability length - in time and in space,
respectively - as:
\begin{eqnarray}
T_{c} & \sim & Q\sigma^{-1} \\
Z_{c} & \sim & gt_{c}^2 \sim Q^2 g \sigma^{-2}
\end{eqnarray}
Note that if $h(z) \sim z^{-b}$, and thus $v(z) \sim z^{2b}$, the thread 
would be stable if $2/7 < b < 1/2$. This could be realized if the 
gravitational field increased as $g(z) \sim z^a$ with $a=4b-1$. The case 
studied above, $b=1/4$, is slightly below the lower limit of stability. 

\subsection{The effect of viscosity: spatially growing modes.}
With finite viscosity, linearization of (\ref{ourmodel_1}) and 
(\ref{ourmodel_2}) 
in the same way as in the previous section equations leads to
\begin{eqnarray}
a_{t}+a_{y}+2W^{-1}W_{y}a&=&q^{-1/2}W^{-3/2}b_{y}\nonumber \\
&+&\frac{q^{-1/2}}{2}W^{-5/2}W_{y}b\label{visc_15} \nonumber \\ 
&+&\gamma W^{-2} \left( a_{yy}+2W^{-1}W_{y}b_{y}\right.\nonumber\\
&+&\left. (W_{yy}-2W^{-2}W^{2}_{y})a \right)  \\
b_{t}+b_{y}+\frac{a_{y}}{2}&=&0 \nonumber
\end{eqnarray}
Perturbing again around the asymptotic state (\ref{v_inert}), $W$ is 
determined by (\ref{str}). Again, we transform to the co-moving frame 
(\ref{comov}), and in the regime $t\gg x$ viscosity drops out - the viscid 
corrections are subdominant. But if we instead assume that $x \gg t$ the 
situation is different. Thus we approximate (\ref{visc_15}) by 
neglecting all explicit time-dependence (coming from $W$). It is thus 
natural to assume
the following behaviour of the amplitudes of perturbation (\cite{Huer},\cite{deLuca}):
\begin{eqnarray}
(a(x,t),b(x,t))=(A(x,s),B(x,s))\exp(st)
\end{eqnarray}  
where $s$ is a real number. 
This leads to the following system of equations:
\begin{eqnarray}
sB&=&-\frac{A_{x}}{2}\\
sA&+&2W^{-1}W_{x}A=q^{-1/2}W^{-3/2}B_{x}\nonumber\\
&+&\frac{q^{-1/2}}{2}W^{-5/2}W_{x}B \nonumber\\
&+&\gamma W^{-2}\left[A_{xx}+2W^{-1}W_{x}B_{x}\right.\nonumber\\
&+&\left. (W_{xx}-2W^{-2}W^{2}_{x})A\right]
\end{eqnarray}
After some manipulations we end up with the single second order
equation:
\begin{eqnarray}
 \label{bound_1}
&& \left(s+\frac{2}{x}+\frac{2\gamma}{x^4}\right)A=
-\frac{q^{-1/2}s^{-1}}{4}x^{-5/2}A_{x}+\\
&+&\left(-\frac{q^{-1/2}s^{-1}}{2}x^{-3/2}+\gamma x^{-2}
-\gamma x^{-3}s^{-1} \right)A_{xx}\nonumber
\end{eqnarray}
Let us now assume that we are looking for the solutions
of (\ref{bound_1}) which grow in the positive
direction of $x$ (see \cite{Keller}). From the
physical point of view this means, that perturbations should
remain finite in the area near the outlet. If
the spatial coordinate satisfies the following condition:
\begin{equation}
\max(2s^{-1},2^{1/4}\gamma ^{1/4}s^{-1/4}) \ll x \ll 4 \gamma ^2 s^2 q
\label{cond_1}
\end{equation} 
the dominant terms in (\ref{bound_1}) are give the simpler equation
\begin{eqnarray}
A_{xx}-\frac{q^{-1/2}\gamma^{-1}s^{-1}}{4}x^{-1/2}A_{x}
-\frac{s}{\gamma}x^2 A=0 \label{bound_2}
\end{eqnarray}
with the WKB-type solution
\begin{eqnarray}
A_{+} \propto \exp\left(\sqrt{\frac{s}{4\gamma}}x^2\right) \label{wkb_sol}
\end{eqnarray}
which is valid for:
\begin{eqnarray}
 \max(\gamma ^{1/4}s^{-1/4},\gamma ^{-1/3}s^{-1/3}q^{-1/3}) \ll x
\label{cond_2}
\end{eqnarray}
We conclude that  viscosity gives rise to a superexponential growth along 
the spatial variable $x$ with a characteristic length $l_{\nu} \sim 
\gamma^{1/4}$, which in dimensional variables becomes
 $L_{\nu} \sim  l_c \gamma^{1/4}=  \sigma^{5/16} g^{-9/16}\nu^{1/4}$.
\par
For the large values of $x$ the viscous effects drop out and 
(\ref{bound_1}) leads to
\begin{equation}
A_{xx}+\frac{1}{2x}A_{x}+2sq^{1/2}x^{3/2}A=0
\end{equation}
which produces slowly decaying WKB-solutions:
\begin{equation}
A_{\pm} \propto x^{-1/4}\exp(\pm i \sqrt{2} |s| q^{1/4} x^{7/4} ) 
\end{equation}
We can conclude, that in the first time instant $t \ll x $ the dynamics
of the system
is governed by spatially growing modes. But the region of validity for 
the "spatial asymptotics" shrinks with time as $x \propto t$, and the 
"temporal
asymptotics"  ($\propto \exp(t^{1/4})$) takes over. For large enough 
$\gamma$, this regime will finally define the break-up of a flow unless it 
have been destroyed already by spatially growing modes (the latter seems 
to happen in the limit of small $\gamma$, e.g. for water).

\section{Discussion.}
The stationary flow of a long  falling viscous thread, with the
asymptotic shape
$h(z) \sim z^{-1/4}$ is unstable as expected from the classical results
for the stability of a fluid cylinder. 
The perturbations grow, however, very slowly, increasing asymptotically
only as $ \exp( {\rm const}  \, t^{1/4})$, where the constant is
 {\it independant of viscosity}.
What is the role of viscosity then? First, with respect to stationary
solution, viscosity defines the structure of the flow near the outlet.
The crossover between viscous and inertial solutions is found to scale
like $\propto \nu^{2/3}$. The less
viscous threads are vulnerable due to spatial instability, since the 
perturbations grow like $\exp(\nu^{-1/2}x^2)$ along the thread. 
For large viscosities, the effects of spatial instability are 
weak, and the inviscid asymptotics will dominate 
the development of the break-up. It is interesting to note, that the final 
instability is so weak, that 
if the gravitational field was growing slowly i.e.  $g(z) \sim z^a$, with 
$1/7 <a <1$, the thread would
become asymptotically stable.
\section*{ACKNOWLEDGEMENTS}
We would like to thank Jens Eggers for helpful advice and several  
important discussions. T. B thanks the Danish Natural Science Research 
Council for support.
\section*{APPENDIX}
\par
In this appendix we show that the explicit dependence on $x$ in 
(\ref{lin1b}) can be neglected when $t \gg |x|$.
Neglecting this variation led to the "local" solution (\ref{one_q1}):
\begin{eqnarray}
\label{basic_solb}
\tilde b_{0}(k,t)&=&t^{-1}\exp(\varepsilon kt^{1/4}) \\
\tilde a_{0}&=&\frac{2i}{k}\tilde b_{0t}
\label{basic_sola}
\end{eqnarray}
where $\varepsilon= \sqrt{8}q^{-1/4}$ and where we define the direct and 
inverse Fourier transforms as
\begin{eqnarray}
\tilde a(k,t)&=&\frac{1}{2\pi}\int a(x,t)\exp(-ikx)dx, \\
a(x,t)&=&\int \tilde a(k,t)\exp(ikx)dk \nonumber
\end{eqnarray}
Let us now include effetcs of non-locality by Fourier transformation of 
(\ref{lin1b})
\begin{eqnarray}
\nonumber
\tilde a_{t} &+& \frac{1}{2\pi}\int{\frac{2a\exp({-ikx})}{x+t}}dx \\
&=&q^{-1/2}\frac{1}{2\pi}\int{\frac{b_{x}\exp({-ikx})}{(x+t)^{3/2}}}dx\nonumber\\
&+&q^{-1/2}\frac{1}{2\pi}\int{\frac{b\exp({-ikx})}{2(x+t)^{5/2}}}dx 
\label{int_k}
\end{eqnarray}

Now we assume that $|x|\ll t$ and expand integral kernels in (\ref{int_k})
in power series of $|x|/t$. First we consider LHS  of (\ref{int_k}):
\begin{eqnarray}
&&\frac{1}{2\pi}\int{\frac{2a\exp({-ikx})}{x+t}}dx= 2t^{-1}\tilde a\nonumber\\
&+&2\sum_{n=1}^{\infty}t^{-n-1}C_{-1}^{n}
\frac{1}{2\pi}\int x^{n}a\exp({-ikx})dx
\end{eqnarray}
We substitute:
\begin{eqnarray}
\frac{1}{2\pi}\int
x^{n}a\exp({-ikx})dx=(i)^{n}\left(\frac{d}{dk}\right)^{n}\tilde a
\end{eqnarray}
and get:
\begin{eqnarray}
&&\frac{1}{2\pi}\int{\frac{2a\exp({-ikx})}{x+t}}dx= 2t^{-1}\tilde a\nonumber\\
&+&2\sum_{n=1}^{\infty}t^{-n-1}C_{-1}^{n}
(i)^{n}\left(\frac{d}{dk}\right)^{n}\tilde a \label{rhs1}
\end{eqnarray}
where $C_{m}^{n}$ is a binomial coefficient.
Let us estimate correction term in the RHS of (\ref{rhs1}) for the local 
solution
 (\ref{basic_solb})-(\ref{basic_sola}). After some algebra we obtain:
\begin{eqnarray}
&&\frac{1}{2\pi}\int{\frac{2a_{0}(t) \exp({ikx})}{x+t}}dx= 2t^{-1}\tilde 
a_{0}(t)\\
&+&\tilde b_{0}(t)
\sum_{n=1}^{\infty}t^{-n}C_{-1}^{n}(i)^{n+1}
\sum_{m=0}^{n}C_{n}^{m}(n-m)! \nonumber\\
&\times&(-1)^{m-1}k^{-1-n+m}\varepsilon^{m} 
t^{\frac{m-7}{4}}
(
k\varepsilon+(m-4)t^{-1/4}
\nonumber
)
\end{eqnarray}
Thus the highest power of $t$ under the inner summation
is $\propto t^{(n-7)/4}$. Taking into account general multiplier
$\propto t^{-n-1}$, in the leading order we have the correction:
$\propto t^{-(3n+7)/4}t^{-1}\exp(\varepsilon k t^{1/4})$. This should be
compared with the main contribution on the rhs. (\ref{int_k}),
i.e. $t^{-3/2}t^{-1}\exp(\varepsilon k t^{1/4})$. Obviously:
\begin{eqnarray}
t^{-(3n+7)/4}< t^{-3/2},\ \ \ t \mapsto \infty, \ \ \ n=1,2,...
\end{eqnarray}
Now we continue with the rhs. of (\ref{int_k}). Applying the same
method, we get:
\begin{eqnarray}
&&\frac{1}{2\pi}\int{\frac{b_{x}\exp({-ikx})}{(x+t)^{3/2}}}dx\nonumber \\
&=&t^{-3/2}\sum_{n=0}^{\infty}t^{-n}C_{-3/2}^{n}(i)^{n+1}
\left(\frac{d}{dk}\right)^{n}(k \tilde b) 
\end{eqnarray}  
For the correction term we get:
\begin{eqnarray}
\propto 
t^{-3/2}\sum_{n=1}^{\infty}t^{-\frac{3n}{4}}C_{-3/2}^{n}(i)^{n+1}\varepsilon^{n-1}
\left[
k\varepsilon+n t^{-1/4}
\right] \tilde b_{0}
\end{eqnarray}
Now the leading order term is $\propto t^{-3n/4}t^{-3/2}\tilde b_{0}(t)$,
which is dominated by the main term 
$\propto t^{-3/2}\tilde b_{0}(t)$ when $t \rightarrow \infty$. As for the 
last
term in the RHS of (\ref{int_k}),
it only produces minor corrections to the main solution even in the
leading order of magnitude. Thus we see
that for $|x|/t \ll 1$ we can completely neglect the effects of
non-locality in (\ref{int_k}).


\begin{references}

\bibitem{chandra} S. Chandrasekhar, {\em Hydrodynamic and Hydromagnetic 
Stability},
(Dover, New York, 1981).

\bibitem{E97}  J. Eggers, Rev. Modern Phys., {\bf{69}}, 865 (1997)

\bibitem{LL} L. D. Landau, and E. M. Lifshitz, {\em Fluid
Mechanics}, (Pergamon, Oxford, 1984).

\bibitem{QJ} N. S. Clarke, Quart. Journ. Mech. and Applied Math.,  
{\bf{XXII}},
247 (1968)

\bibitem{ReAcT} A. Kaye and D. G. Vale, Rheologica Acta {\bf{8}},
1 (1969)

\bibitem{Tuck} E. O. Tuck, J. Fluid. Mech. {\bf{76}},
625 (1976)

\bibitem{Gee1} J. Geer, Phys. Fluids {\bf{20}},
1613 (1977)

\bibitem{Gee2} J. Geer, Phys. Fluids {\bf{20}},
1622 (1977)

\bibitem{G&S} J. Geer and J. C. Strikwerda J. Fluid Mech. {\bf{135}},
155 (1983)

\bibitem{Huer} P. Huerre and P. A. Monkewitz Annu. Rev. Fluid Mech. {\bf{22}},
473 (1990)

\bibitem{deLuca} L. de Luca and M. Costa, J. Fluid Mech. {\bf{331}},
127 (1997) 

\bibitem{Keller} J. B. Keller, S. I. Rubinov and Y. O. Tu
 Phys. Fluids {\bf{16}}, 2052 (1973) 

\end{references}
\end{document}